\documentclass{aastex}
\everymath{\rm}
\everydisplay{\sf}
\received{}
\revised{}
\accepted{}
\cpright{}{}\journalid{}{}
\articleid{}{}
\paperid{}
\slugcomment{To appear in {\it PASP}}
\shortauthors{Kwitter \& Henry}
\shorttitle{Testing S, Cl, \& Ar Nucleosynthesis in PNe}
\begin{document}
\title{Sulfur, Chlorine, and Argon Abundances in Planetary
Nebulae. III: Observations and Results for a Final Sample}

\author{K.B. Kwitter\footnote{Visiting Astronomer, Kitt Peak National
Observatory, National Optical Astronomy Observatories,
which is operated by the Association of Universities for Research in
Astronomy, Inc. (AURA) under cooperative agreement with the
National Science Foundation}}
 
\affil{Department of Astronomy, Williams College, Williamstown, MA
01267; kkwitter@williams.edu}
 
\author {R.B.C. Henry$^{1,}$\footnote{Visiting Astronomer, Cerro Tololo Interamerican
Observatory, National Optical Astronomy Observatories,
which is operated by the Association of Universities for Research in
Astronomy, Inc. (AURA) under cooperative agreement with the
National Science Foundation.}}
 
\affil{Department of Physics \& Astronomy, University of Oklahoma,
Norman, OK 73019; henry@mail.nhn.ou.edu} 

\and

\author{J.B. Milingo$^{1,2}$}

\affil{Department of Physics, Gettysburg College, Gettysburg, PA
17325; jmilingo@gettysburg.edu}

\begin{abstract}

This paper is the fourth in a series whose purpose is to study the
interstellar abundances of sulfur, chlorine, and argon in the Galaxy
using a sample of 86 planetary nebulae. Here we present new
high-quality spectrophotometric observations of 20 Galactic planetary
nebulae with spectral coverage from 3700-9600~{\AA}. A major feature
of our observations throughout the entire study has been the inclusion
of the near-infrared lines of [S~III] $\lambda\lambda$9069,9532, which
allows us to calculate accurate S$^{+2}$ abundances and to either
improve upon or convincingly confirm results of earlier sulfur
abundance studies. For each of the 20 objects here we calculate ratios
of S/O, Cl/O, and Ar/O and find average values of
S/O=1.1E-2$\pm$1.1E-2, Cl/O=4.2E-4$\pm$5.3E-4, and
Ar/O=5.7E-3$\pm$4.3E-3. For six objects we are able to compare
abundances of S$^{+3}$ calculated directly from available [S~IV]
10.5$\mu$ measurements with those inferred indirectly from the values
of the ionization correction factors for sulfur. In the final paper of
the series, we will compile results from all 86 objects, search for
and evaluate trends, and use chemical evolution models to interpret
our results.

\end{abstract}


\clearpage

\section{INTRODUCTION}

This is the fourth paper in a project to study abundances in planetary
nebulae (PNe), highlighting the elements sulfur, chlorine, and argon.
The motivation to focus on these three elements has been the
realization that their abundances are unaffected by nucleosynthesis in
the progenitors of PNe. In contrast to elements like carbon, nitrogen,
and oxygen, whose current nebular abundances can be profoundly
different from the those in the original star, the sulfur, chlorine
and argon abundances we measure now in a nebula are representative of
the original composition of the progenitor star. In this way, PNe can
provide information on both current and past chemical composition of
the interstellar medium in the Galaxy, and can be used to evaluate
theoretical yield predictions from stellar evolution models. 

The three preceding papers in this series (Kwitter \& Henry 2001
[Paper~I], Milingo et al. 2001 [Paper~IIA], and Milingo, Henry, \&
Kwitter 2001 [Paper~IIB]) have reported spectrophotometry and
abundance analyses of a total of 56 primarily type~II (disk) PNe in
our Galaxy. In the present paper we present new spectrophotometry and
abundances for 20 more objects. In the final paper, we will also include
results from 10 PNe originally observed for another project, bringing
the total in our sample to 86.

In \S 2 we describe the observations and data, and in \S 3 we present
our results; a summary is given in \S 4.

\section{DATA}

Table 1 lists the objects discussed in this paper.  Column~1 gives the
object name, column~2 the angular size, column~3 the slit offset, if any,
and columns~4 and 5 the blue and red exposure times,
respectively. Column~6 contains either K or C denoting
optical/near-infrared spectrophometry obtained at Kitt Peak or Cerro
Tololo, respectively. Six PNe have available ISO SWS observations that
include the [S~IV] 10.5$\mu$ line, which we discuss in \S 2.2; these
objects are listed first in Table~1 and the ISO exposure times are
given in the last column.

\subsection{Optical and Near-Infrared Observations}

Observations at CTIO were obtained in March 1997 using the 1.5m
telescope and cassegrain spectrograph with Loral 1K CCD.  The 1200 x
800 Loral 1K CCD has 15$\mu$ pixels.  We used a 5$\arcsec$ x
320$\arcsec$ extended slit in the E-W direction.  Perpendicular to
dispersion the scale was 1.3$\arcsec$/pixel.  Gratings \#22 and \#9
were used to obtain extended spectral coverage from 3600-9600 {\AA}
with overlap in the H$\alpha$ region.  Both gratings have nominal
wavelength dispersions of 2.8 \AA/pixel and 8.6 {\AA}~FWHM resolution.

Data for objects observed at KPNO were obtained in May 1996,
December 1996, or June 1999 with the Goldcam CCD spectrometer at the
2.1m telescope. The chip was a Ford 3K $\times$ 1K CCD with 15$\mu$
pixels. We used a slit that was 5$\arcsec$ wide and extended
285$\arcsec$ in the E-W direction, with a spatial scale of
0$\farcs$78/pixel. The 3700-9600\AA\ range, overlapping coverage from
$\sim$5750 - 6750\AA, was covered with two gratings. For the blue, we
used $\#$240 and a WG345 order-separation filter; wavelength
dispersion was 1.5 \AA/pixel ($\sim$8 \AA\ FWHM resolution). For the
red we used grating $\#$58 with an OG530 order-separation filter,
yielding 1.9 \AA/pixel ($\sim$10 \AA\ FWHM resolution).

Both CTIO and KPNO CCD's produced fringing at wavelengths beyond
$\sim$7500 \AA.  Assuming less than ideal fringe removal via dome
flats, the fringe amplitudes range from roughly $\pm$1\% at 7500 \AA,
to $\pm$9.8\% at 9500 \AA, the longest wavelength we measure. We note
this contribution to the uncertainty in our line intensities measured
at wavelengths longer than 7500 \AA. 

Many of these PNe are relatively small in angular size; in those cases, we
placed the spectrograph slit on the brightest part of the nebula as
seen on the acquisition screen, avoiding the central star if it was
visible. We obtained the usual bias and twilight flat-field frames
each night, along with HeNeAr comparison spectra for wavelength
calibration and standard star spectra for sensitivity calibration. The
original two-dimensional spectra were reduced and calibrated using
standard long-slit spectrum reduction methods in IRAF\footnote{IRAF is
distributed by the National Optical Astronomy Observatories, which is
operated by the Association of Universities for Research in Astronomy,
Inc. (AURA) under cooperative agreement with the National Science
Foundation.}.  One-dimensional spectra were extracted from the
original two-dimensional images interactively using the {\it kpnoslit}
package. Line fluxes were measured with {\it splot}.

Our measured line strengths are tabulated in Tables~2A-C. For each
line identification in the first column we list the relative
extinction factor $f(\lambda)$ followed by two columns for each object
containing the observed [F($\lambda$)] and dereddened [I($\lambda$)]
line strengths normalized to H$\beta$=100. Observed fluxes were
dereddened using the extinction curve of Savage \& Mathis (1979),
where each dereddened intensity was obtained by multiplying the
observed flux by the factor 10$^{cf_{\lambda}}$, where $c$ is the
logarithmic extinction factor and $f_{\lambda}$ is the reddening
coefficient. We assumed H$\alpha$/H$\beta$=2.86 (Hummer \& Storey
1987). Values for the logarithmic extinction factor $c$ and
log~F$_{H\beta}$ in erg cm$^{-2}$ s$^{-1}$, the H$\beta$ flux as
measured through the slit are found at or near the bottom of each
table. Estimated uncertainties for line strengths are represented
using colons, as defined in the table footnotes.

To test the accuracy of our extinction factor, $c$, we determined
separate values using the Paschen~10 and Paschen~8 lines at 9014{\AA}
and 9546{\AA}, respectively. Ideally, these $c$ values should match
the one inferred using H$\alpha$, i.e. the one listed in the line
strength tables. This comparison is shown in Table~3A, where for each
object we list the three values of $c$. Recombination data required
for this exercise were taken from Osterbrock (1989, Table~4.4). These
values are further compared in Fig.~1, where we plot c(P10) or c(P8)
versus c(H$\alpha$) and include a straight line to show the ideal
one-to-one relation. Overall, the trend is as anticipated, but with a
typical difference of $\sim$0.15 between c(H$\alpha$) and either
c(P10) or c(P8).

As a check on the quality of the data and measuring accuracy, we list
in Table~3B our measurements for several line ratios predicted by
theory to have constant values. These ratios are identified in the
table footnote. The last two lines of the table give our observed mean
and the theoretical (expected) value. With the exception of the
[Ne~III] ratio, agreement is good. 

In {\S}3 we employ the dereddened intensities to determine electron
temperatures, densities, and abundances.

\subsection{ISO Data}
We were able to obtain pre-publication estimates of ISO SWS fluxes for
[S~IV] 10.5$\mu$ and Br$\alpha$ 4.05$\mu$ for the six PNe in
Table~2A (Barlow \& Liu, private communication; Beintema \& Salas,
private communication). These observed fluxes are listed near the
bottom of Table~2A. We joined them to the optical spectrum in the
following way.  To correct the observed ratio I(10.5$\mu$)/Br$\alpha$
for differential interstellar extinction, we used the infrared
extinction law, A($\lambda$)/A($V$) given by Rieke \& Lebofsky (1985):
for Br$\alpha$ and 10.5$\mu$ the values are 0.04 and 0.08,
respectively. Combined with our value for {\it c} determined from the
Balmer lines, we took {\it c} = 1.41E(B-V), and {\bf R}, the ratio of
total to selective extinction = 3.1, to calculate A($V$). The
resulting values for A(Br$\alpha$) and A(10.5$\mu$) are shown near the
bottom of Table~2A. For an electron temperature of 10,000 K, and a
density of 1000 cm$^{-3}$, the predicted ratio of Br$\alpha$/H$\gamma$
= 0.17, and H$\gamma$/H$\beta$ = 0.46 (Hummer \& Storey 1987). We
multiplied the reddening-corrected [S~IV]/Br$\alpha$ ratios by 0.17 x
0.46 = 0.078 to yield I($\lambda$) values for [S~IV], which are given
(relative to H$\beta$=100) in Table~2A.

The fluxes for [Ar~III] 8.99$\mu$ were also available for this set of
six PNe (Barlow \& Liu, private communication; Beintema \& Salas,
private communication). These were corrected for extinction
[A($\lambda$)/A($V$) for 8.99$\mu$ = 0.07; Reike \& Lebofsky 1985] and
converted to a ratio with H$\beta$ in a manner analogous with the
[S~IV] 10.5$\mu$ line described above. These data are also presented
near the bottom of Table~2A.

\section{RESULTS}

Electron temperatures, densities, as well as ion and elemental
abundances were calculated using exact routines employed in Papers~I
and IIB. The reader is referred to Paper~I, where our methods are
described in detail. To summarize the methods, temperatures and
densities were computed in the standard fashion, using forbidden line
ratios which are sensitive to the respective quantity. For ion
abundances, we employed a program which utilizes a 5-level atom
equilibrium calculation to compute abundances from forbidden lines of
heavy elements. Ions of H and He were determined from permitted line
strengths using effective recombination coefficients. The sum of
abundances of observed ions for an element was then corrected for
unseen ions through the use of ionization correction factors (ICF)
described in detail in Paper~I to obtain total elemental abundances.

Temperatures, densities, ion abundances, and ICFs are listed by object
in Tables~4A-C. Uncertainties are described in the table
footnotes. Temperatures which appear to be anomalously high are
enclosed in parentheses. For K648 and NGC~2242, we were unable to
determine a [S~II] electron density directly from our data, because
strengths for one or both of the 6716~{\AA} and 6732~{\AA} lines was
undetected. In this case, densities reported in earlier papers were
adopted, as indicated in footnotes to the relevant tables. A major
feature of our work in this series of papers is to use the
near-infrared (NIR) forbidden lines of [S~III] at
$\lambda\lambda$9069,9532 along with $\lambda$6312 to derive S$^{+2}$
abundances using a [S~III] temperature where possible. In a few cases
where this temperature differed by more than 5,000~K from the [O~III]
temperature, we adopted the [N~II] temperature as more representative
of the S$^{+2}$ zone. For NGC~1535 (Table~4C), however, the [N~II]
value appeared to be suspiciously high, so in this case we used the
[O~III] temperature.

As a check on our S$^{+2}$ measurements, we also calculated the
abundance of this ion using the [S~III] $\lambda$6312 line and
(usually) the [N~II] temperature. These results are included in
Table~4A-C. Although the $\lambda$6312 line is much weaker than the
nebular [S~III] lines in the NIR, it is generally more accessible
spectroscopically, and thus has been used extensively to calculate
S$^{+2}$ abundances (cf. Kingsburgh \& Barlow 1994). We plot the two
abundances against each other in Fig.~2 and include a straight line to
show the ideal one-to-one correspondence. As we have also found in
Papers~I and IIB, the NIR lines tend to yield lower abundances for
this ion. This could result from the fact that when [S~III]
$\lambda$6312 is used, the [S~III] temperature is generally
unavailable, and one is forced to employ either the [O~III] or [N~II]
temperature. Since generally we found these temperatures to be
somewhat lower than the [S~III] temperature in the cases where all
three could be determined, that would explain the higher S$^{+2}$
abundances when the [S~III] temperature was unavailable.

As a first-order check on our sulfur ICF, we were able to use the ISO
data described above for the six PNe in Table~4A to calculate the
ionic abundance of S$^{+3}$ from the [S~IV] 10.5$\mu$ line, using
collision strengths and transition rates from Mendoza (1983). These
results, labelled ``IR'' are compared in Table~5 with S$^{+3}$
abundances inferred by assuming that $S^{+3} = (S^{+2} + S^+) \times
(ICF-1)$, labelled ``ICF.'' The fourth column in the table lists the
ratio of ICF/IR abundances. If the methods agreed perfectly, then
values in this column would be unity; clearly this is not the
case. Some of the variance is surely due to the uncertainties in the
values in Table~5, but this is likely not the whole explanation. It is
always risky to compare observations of an extended PN obtained
through apertures that differ in size and/or placement. We note that
there are significant differences in the nebular area sampled by the
two kinds of observations presented here: the ISO SWS aperture is
roughly 20$\arcsec$ x 33$\arcsec$ (with some dependence on
wavelength), while our observations were made through a
5$\arcsec$-wide slit. Except for Hb~12, which is notoriously
inhomogeneous (Hyung \& Aller 1996), and where the (uncertain) ICF
value of S$^{+3}$/H$^+$ is about eight times that indicated by the
(very uncertain) 10.5$\mu$ line, the ratios for the other PNe, some
also fairly uncertain, range from 0.22 to 0.88. So, while not
as close as one might wish, the values in Table~5 are not excessively
discordant, considering the uncertainties in the ISO calibrations, in the
conversion of infrared fluxes to intensity ratios with H$\beta$, and in the
different nebular areas sampled.

As a second comparison with ISO data, we used fluxes for the [Ar~III]
8.99$\mu$ line. In Table~4A we list the Ar$^{+2}$/H$^+$ abundances
calculated from the 8.99$\mu$ and 7135~{\AA} lines; it can be seen
that all pairs of values agree to within a factor of two, and for four
of the six objects, agreement is within 20\%. This agreement for
abundances calculated separately from observed fluxes of two lines of
the same ion, is heartening. It may be that measurements of [Ar~III]
are less sensitive than [S~IV] to differing slit sizes and
positions. In conclusion, while agreement between abundances of
S$^{+3}$ and Ar$^{+2}$ inferred from ICF predictions and derived
directly from observations is encouraging, uncertainties introduced by
the differences in aperture size make it difficult to make a more
detailed assessment.

Elemental abundances for all 20 objects are listed in Tables~6A-C. The
last two columns of each table give solar and Orion Nebula abundances
for comparison, where these have been taken from Grevesse et
al. (1996) and Esteban et al. (1998), respectively. In all cases the
sulfur abundance is based upon the use of S$^{+}$ and S$^{+2}$, along
with the relevant ICF. In Table~6A we also list the final sulfur
abundance determined using the S$^{+3}$ reported in Table~4A for
comparison. Clearly, the discrepancies associated with S$^{+3}$ that
were noted when the ion abundances were considered above have been
suppressed because of the dominant influence of the S$^{+2}$ abundance.

The ratios of S/O, Cl/O and Ar/O are plotted against O/H in
Fig.~3. Results from this paper are shown with filled circles, while
results from Paper~I are shown with open circles. In each panel the
sun's position is indicated with a star, while that of Orion is shown
with an `X'. Typical uncertainties are indicated and explained in the
caption. Clearly there is good agreement between samples. We
also point out the anomalously high values for S/O and Ar/O for Hb12,
the PN with the lowest value for O/H. It is possible that our estimate
of O/H is too low, although we point out that Cl/O appears consistent
with the remainder of the sample.

Finally, Table 7 lists unweighted averages and standard deviations for
the sample in this paper as well as the samples in Papers~I and
IIB. For comparison, we also give the averages found by Kingsburgh \&
Barlow (1994; KB) and Aller \& Keyes (1987; AK) as well as solar and
Orion values. We note that all of our averages compare favorably with
these other results, considering all of the uncertainties.

As we have seen previously, the ratios S/O, Cl/O, and Ar/O appear to
be constant over the metallicity range as gauged by O/H. This is
consistent with the idea that these four elements are formed by
stars belonging to the same mass range. In our final paper in the
series we plan to explore this idea in detail, employing chemical
evolution models in order to evaluate current predictions of stellar
nucleosynthesis.

\section{SUMMARY}

In this, the fourth paper in a series investigating the abundances of
S, Cl, and Ar in planetary nebulae and the Galactic interstellar
medium, we report on spectrophotometric observations of 20 Galactic
PNe, where our spectral coverage extended from 3700-9600~{\AA}. We
also calculate electron temperatures and densities, as well as ion and
element abundances for each object. We find average values of
S/O=1.1E-2$\pm$1.1E-2, Cl/O=4.2E-4$\pm$5.3E-4, and
Ar/O=5.7E-3$\pm$4.3E-3 for our sample. These numbers agree very well
with our results in the previous papers in the series and with
determinations made by others. One of the major features of our
current work is to use the NIR lines of [S~III]
$\lambda\lambda$9069,9532 along with [S~III] temperatures to calculate
sulfur abundances. This is the first time these lines have been used
to determine sulfur abundances in such a large PN sample.

In the next (final) paper we will compile results from all papers
in this series and search for trends among the data. In addition, we
plan to apply chemical evolution models to our data in an attempt to
evaluate the quality of published stellar yields for O, S, Cl, and Ar,
as well as to assess the role of Type~Ia supernovae in the cosmic
buildup of these four elements.

\acknowledgments

We thank the TACs at KPNO \& CTIO for granting us observing time, to
the local staff there for their assistance, and to the IRAF staff for
their ready answers. We are extremely grateful to M.J. Barlow and
X.-W.Liu and to D. Beintema and J.B. Salas for providing the ISO SWS
line fluxes. Our research is supported by NSF grant AST-9819123.

\clearpage



\clearpage

\begin{figure}
\figurenum{1}
\plotone{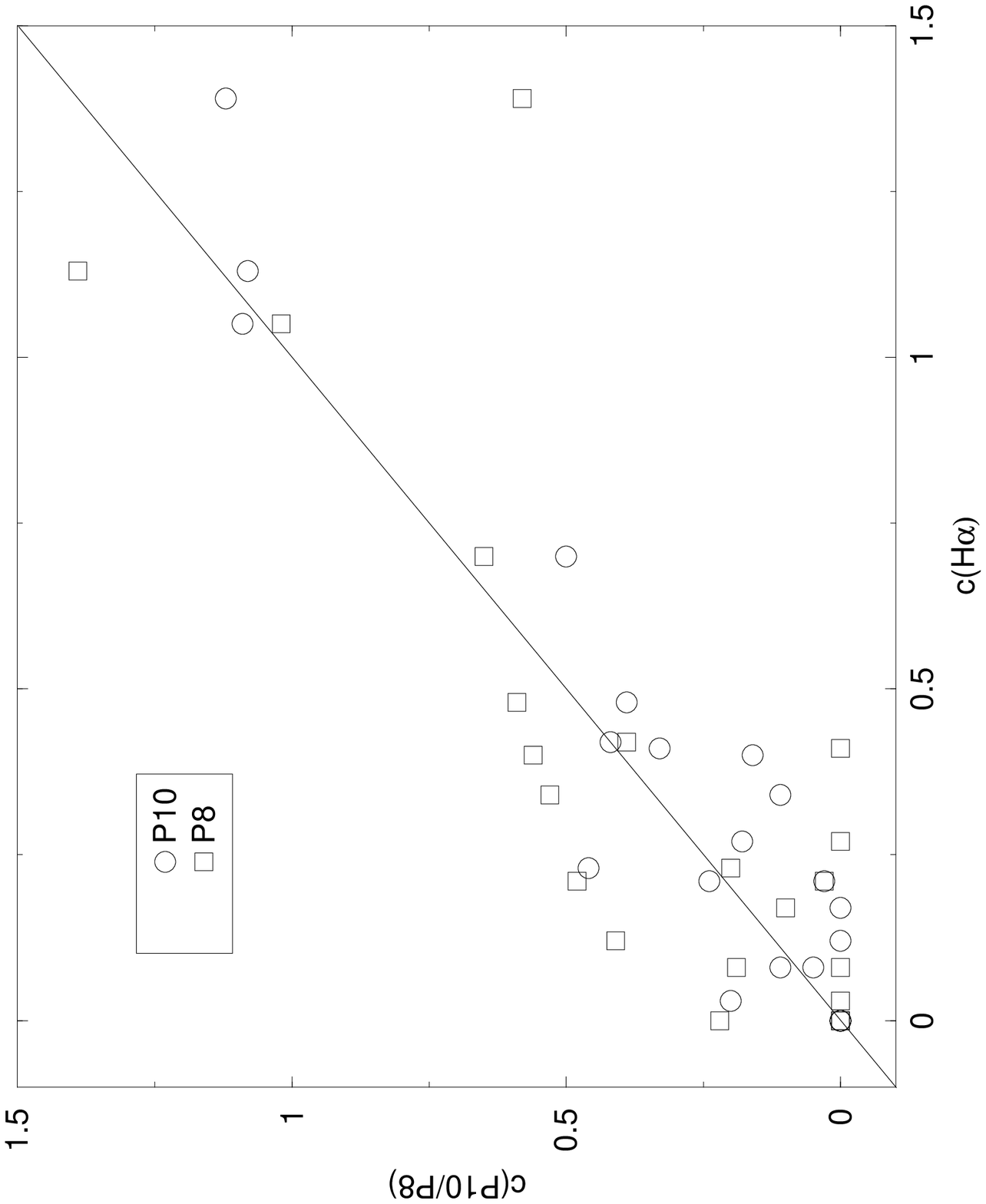}
\caption{A comparison of logarithmic extinction {\it c} as determined
using the Paschen 8 and 10 lines versus the value inferred from using
H$\alpha$. The solid line shows the track for a one-to-one
correspondence.}
\end{figure}

\begin{figure}
\figurenum{2}
\plotone{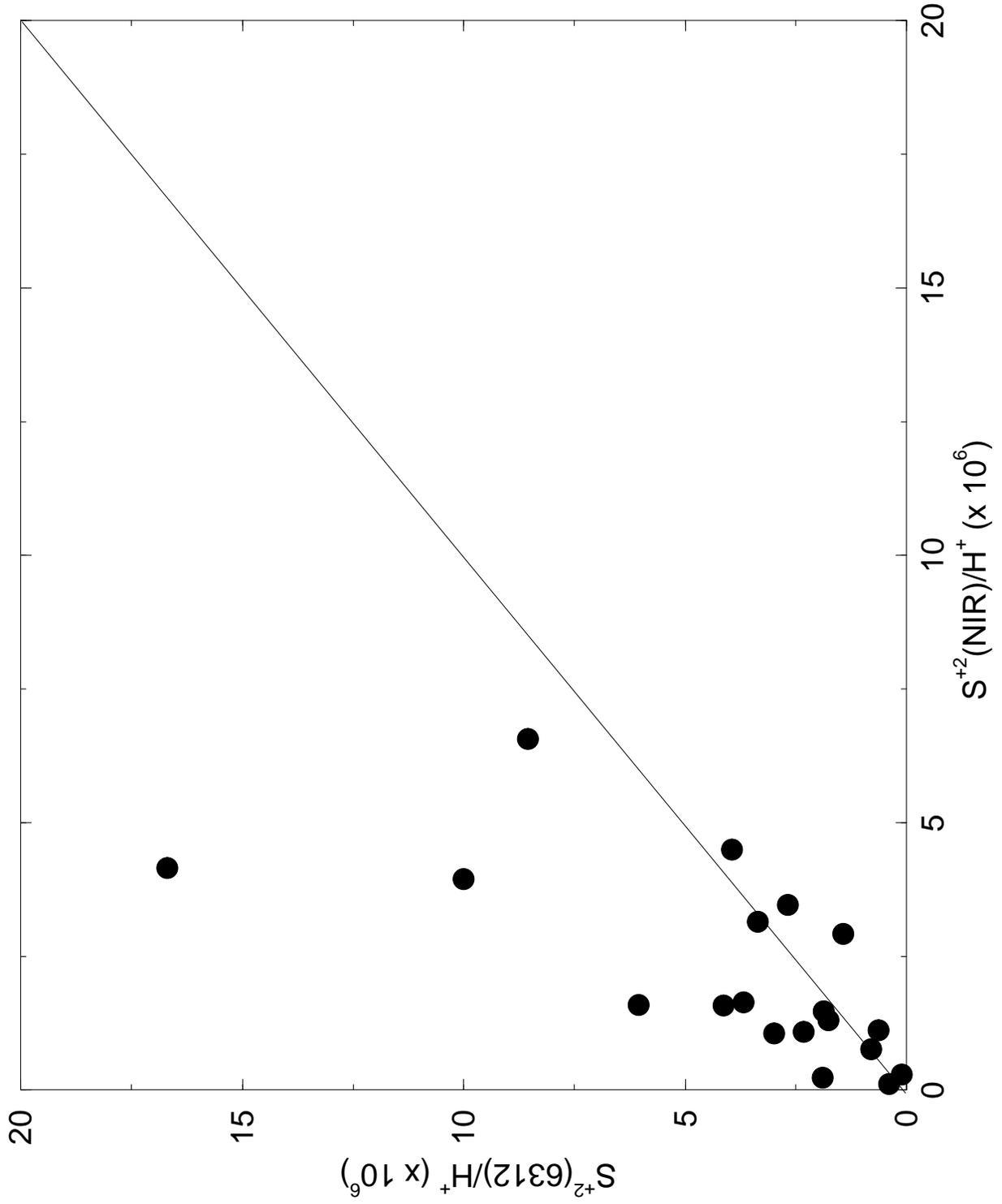}
\caption{Comparison of S$^{+2}$/H$^+$ for S abundances computed using
the 6312{\AA} emission line along with the [N~II] temperature
(ordinate) and the NIR emission lines along with the [S~III]
temperature (abscissa). The solid line shows the track of a one-to-one
correspondence.}
\end{figure}

\begin{figure}
\figurenum{3}
\plotone{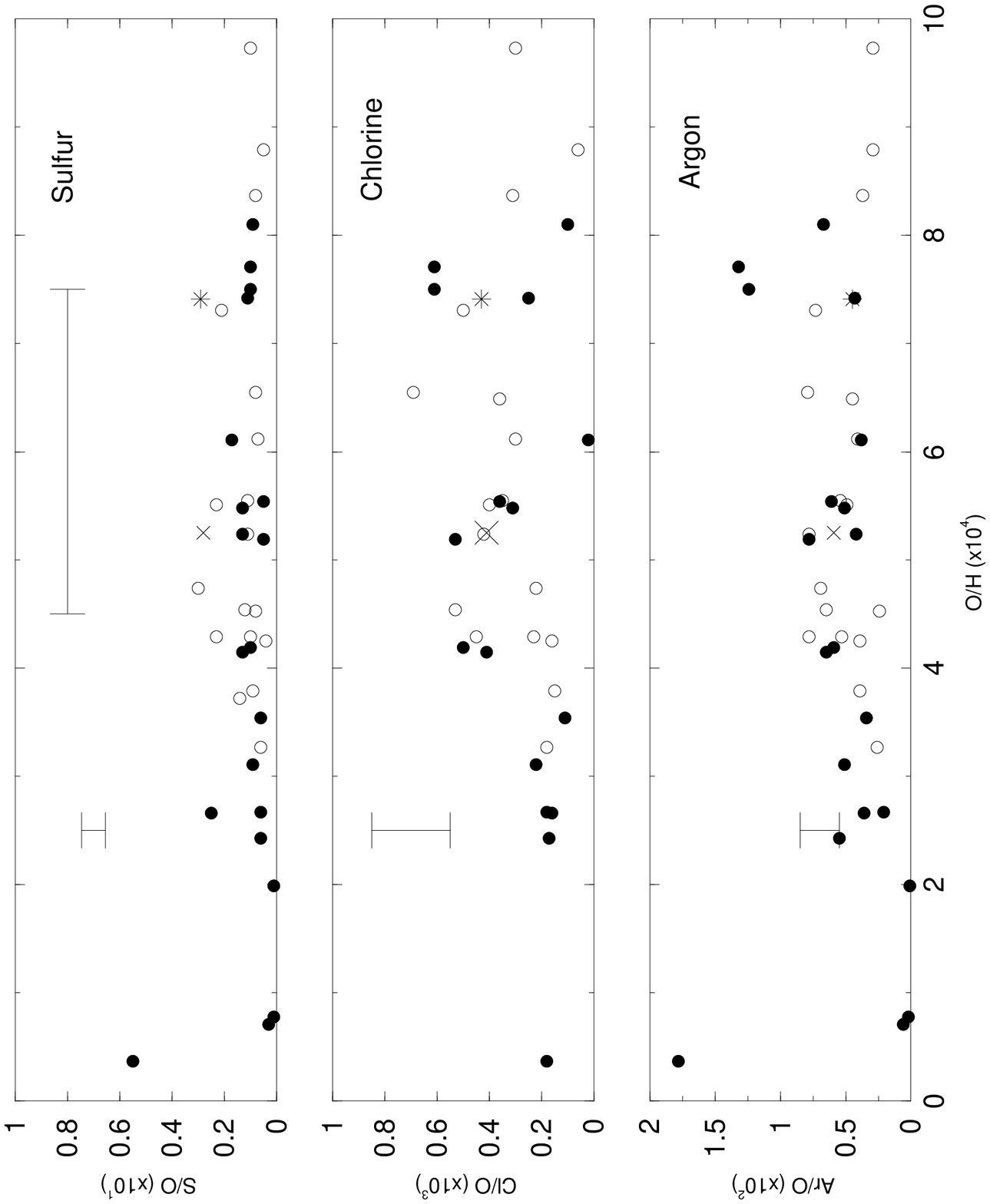}
\caption{Top: S/O (x 10$^1$) versus O/H (x 10$^4$), where filled
circles are ratios determined in this paper, open circles are ratios from 
Paper~1. The position of the sun
(Grevesse et al. 1996) and the Orion Nebula (Esteban et al. 1998) are
indicated with a star and an X, respectively. Middle: Same as top but
for Cl/O (x 10$^3$). Bottom: Same as top but for Ar/O (x
10$^2$). Ordinate uncertainties are shown with error bars in each
panel, while the horizontal error bar in the top panel shows the O/H
uncertainty for all three panels.}
\end{figure}


\clearpage

\begin{thebibliography}{}  % leave the {} just like it is.

\bibitem[]{} Aller, L.H., \& Keyes, C.D. 1987, \apjs, 65, 405

\bibitem[]{} Esteban, C., Peimbert, M., Torres-Peimbert, S., \& Escalante, V. 1998, \mnras, 295, 401

\bibitem[]{} Grevesse, N., Noels, A., \& Sauval, A.J. 1996, in ASP Conf. Ser. 99, Cosmic Abundances, ed. S.S. Holt \& G. Sonneborn (San Francisco: ASP), 117

\bibitem[]{} Henry, R.B.C., Kwitter, K.B., \& Bates, J.A. 2000, \apj, 531, 928

\bibitem[]{} Howard, J.W., Henry, R.B.C., \& McCartney, S. 1997, \mnras, 284, 465

\bibitem[]{} Hummer, D.G., \& Storey, P.J. 1987, \mnras, 224, 801

\bibitem[]{}Hyung, S., \& Aller, L.H. 1996, \mnras, 278, 551

\bibitem[]{} Kingsburgh, R.L., \& Barlow, M.J. 1994, \mnras, 271, 257

\bibitem[]{}Kwitter, K.B., \& Henry, R.B.C. 1998, \apj, 493, 247

\bibitem[]{} Kwitter, K.B., \& Henry, R.B.C. 2001, \apj, 562, 804 (Paper~I)

\bibitem[]{} Mendoza, C. 1983, in IAU Symp. 103, Planetary Nebulae, ed. D.R. Flower (Dordrecht: Reidel), 143

\bibitem[]{} Milingo, J.B., Kwitter, K.B., Henry, R.B.C., \& Cohen, R.E. 2002, \apjs, 138, 279 (Paper~IIA) 

\bibitem[]{} Milingo, J.B., Henry, R.B.C., \& Kwitter, K.B. 2002, \apjs, 138, 285 (Paper~IIB)

\bibitem[]{} Osterbrock, D.E. 1989, {\it Astrophysics of Gaseous Nebulae and Active Galactic Nuclei}, (Mill Valley, CA: University Science Books)  

\bibitem[]{} Rieke, G.H., \& Lebofsky, M.J. 1985, \apj, 288, 618

\bibitem[]{} Savage, B.D., \& Mathis, J.S. 1979, \araa, 17, 73
 
\end{thebibliography}
\end{document}